\documentclass[aps,
showpacs,tightenlines,fleqn]{revtex4}
\usepackage{graphics}

\input epsf                                

\begin{document}
\bibliographystyle{apsrev}

\newcommand{\half}{\frac{1}{2}}
\newcommand{\D}{\mbox{D}}
\newcommand{\curl}{\mbox{curl}\,}
\newcommand{\ep}{\varepsilon}
\newcommand{\lleq}{\lower0.9ex\hbox{ $\buildrel < \over \sim$} ~}
\newcommand{\ggeq}{\lower0.9ex\hbox{ $\buildrel > \over \sim$} ~}
\newcommand{\tr}{{\rm tr}\, }

\newcommand{\be}{\begin{equation}}
\newcommand{\ee}{\end{equation}}
\newcommand{\bea}{\begin{eqnarray}}
\newcommand{\eea}{\end{eqnarray}}
\newcommand{\beaa}{\begin{eqnarray*}}
\newcommand{\eeaa}{\end{eqnarray*}}
\newcommand{\Lhat}{\widehat{\mathcal{L}}}
\newcommand{\nn}{\nonumber \\}
\newcommand{\e}{{\rm e}}

\title{\Large Phantom scalar dark energy as modified gravity: \\
understanding the origin of the Big Rip singularity}

\author{F.~Briscese$\,^{{a}}$,
E.~Elizalde$\,^{(b)}$\footnote{elizalde@ieec.uab.es},
S.~Nojiri$\,^{(c)}$\footnote{nojiri@phys.nagoya-u.ac.jp}, and
S.~D.~Odintsov$\,^{(b,d)}$\footnote{odintsov@ieec.uab.es, also at TSPU, Tomsk}}
\affiliation{
$^{(a)}$\, Dipartimento di Modelli e Metodi Matematici, Universit\`a di Roma I,\\
Via A. Scarpa 16, I-00161, Roma, Italy \\
$^{(b)}$\,  Institut de Ci\`encies de l'Espai (IEEC-CSIC) \\
Campus UAB, Facultat de Ci\`encies, Torre C5-Parell-2a pl \\
E-08193 Bellaterra (Barcelona) Spain \\
$^{(c)}$\,Department of Physics, Nagoya University, Nagoya 464-8602, Japan \\
$^{(d)}$\,Instituci\`{o} Catalana de Recerca i Estudis Avan\c{c}ats
(ICREA), Barcelona, Spain \\
}

\begin{abstract}

It is shown that phantom scalar models can be mapped into a
mathematically equivalent, modified $F(R)$ gravity, which turns out
to be complex, in general. Only for even scalar potentials is the
ensuing modified gravity real. It is also demonstrated that, even
in this case, modified gravity becomes complex at the region where
the original phantom dark energy theory develops a Big Rip singularity.
A number of explicit examples are presented which show that these two
theories are not completely equivalent, from the physical viewpoint.
This basically owes to the fact that the physical metric in both
theories differ in a time-dependent conformal factor. As a result, an
FRW accelerating solution, or FRW instanton, in the scalar-tensor
theory may look as a decelerating FRW solution, or a non-instantonic
one, in the corresponding modified gravity theory.

\end{abstract}


\maketitle

\tolerance=5000

\section{Introduction}

The explanation of the origin of dark energy and the precise
description of the cosmological structure of the currently accelerating
universe are fundamental challenges of modern cosmology. A good
amount of observational data indicate quite clearly that the
present universe may already be, or may soon enter, in a so-called
phantom or superacceleration era, with an effective equation
of state parameter $w$ slightly less than $-1$ (for a recent
discussion on phantom-favoring observational data, see \cite{data}).
The simplest possibility to realize this phantom dark energy era is
based on the introduction of a phantom scalar, i.e. a
scalar field with negative kinetic energy (for a recent discussion of
scalar phantom cosmology, see \cite{scalar} and refs. therein). The
fundamental property of such a phantom field in the accelerating
FRW universe is the appearance of a finite-time future
singularity (Big Rip) \cite{brett} of the scale factor. Moreover,
a phantom scalar with negative kinetic energy leads to a number
of instabilities and it is unwanted from a physical point of view.

In a situation like that, it is quite natural to search for other
theories, without negative kinetic energy, which may also lead in a
quite natural way to an effective phantom era. A rather straightforward
possibility is modified gravity (for a recent review, see \cite{art2}),
where indeed an effective phantom phase can be realized without a
scalar phantom. It seems clear that the same phantom era can be
alternatively described by modified gravity, by a phantom with some
specific scalar potential, by a phantom-like ideal fluid, etc. Hence,
it is important to investigate the relation between phantom scalar
models and modified gravity, with the final aim to clarify what the
different properties of both theories are.

It is known already that $F(R)$ modified gravity can be always
presented under the mathematically-equivalent form of a (canonical)
scalar-tensor theory, but this could not be proven for the scalar
phantom one. In the present letter we will show that any scalar
phantom theory can be always represented as a modified $F(R)$ gravity
which is, generally speaking, complex but that for some potentials
may be real. The class of phantom theories leading to real modified
gravities will be investigated in some detail.

Before going on, it must be pointed out that, in spite of the
mathematical equivalence of these two theories, they are
not physically equivalent. This is due to the fact that the physical
metric ---which is to be fitted against the observational data--- is
different in both theories, owing to the appearance of a conformal
factor in the process of transforming one theory into the other. As
a result, an FRW instanton in one version is not necessarily an
instanton in the (mathematically) equivalent theory, or an
accelerating FRW cosmological solution in the scalar-tensor theory
may result into a decelerating FRW universe, in the corresponding
version of modified gravity. One must be very careful in analyzing
all these possibilities. Furthermore, the origin of the Big Rip
singularity will be hereby clarified, following the mapping of the
phantom phase into real, modified gravity: the phantom Big Rip will
precisely correspond to the region of modified gravity where it
becomes complex.

\section{The phantom scalar as (complex) modified gravity}

Using a complex conformal transformation, we will show in this
section how the phantom scalar can be represented as an equivalent
modified gravity theory. The starting action for the scalar-tensor
theory will be given by
\be
\label{F1}
S = \int d^4 x \, \sqrt{-g'} \left\{\frac{R'}{2\kappa^2} \mp
\frac{1}{2} \partial_{\mu} \varphi \partial^{\mu} \varphi  - V(\varphi)\right\} \ ,
\ee
where $ V(\varphi)$ is a potential of the scalar field $\varphi$ and
$R'$ is the scalar curvature corresponding to the metric tensor
$g_{\mu \nu}$. In the above expression, the sign
in front of the kinetic term is $+$ ($-$) for the case that
the scalar field $\varphi$ is a phantom (canonical scalar). In
\cite{art1} it has been shown that, in the non-phantom case, the
scalar-tensor theory can be mapped ---using a conformal transformation
of the metric tensor--- to modified gravity (for a general review of
this procedure, see \cite{art2}). The relation between the two
theories has been  investigated with care. Mathematically, the
two are equivalent but physically there are certain non-equivalency
issues \cite{art1,thomas,art3}. They are related with the fact that
the physical metric that has to fit the observational data in the two
theories is different (due to the conformal factor). For instance,
some accelerating FRW universe solution of the scalar-tensor theory
may well correspond to a decelerating FRW universe in the equivalent,
modified gravity formulation (see some examples in \cite{art3}).
Under these circumstances, if it turns out that the cosmological
parameters are well fitted from the ones of the scalar-tensor theory,
or either from those of modified gravity, it is this well behaved
corresponding theory the one which will better describe our current
accelerating universe. Note also that there was recent discussion
\cite{chiba} indicating that some versions of $F(R)$ gravity may have
problems with Solar System tests. We will not discuss different points of
view on this problem here.

A nontrivial problem occurs when a conformal transformation is used
for the phantom case. Indeed, a real conformal transformation on the
metric tensor of the type used in \cite{art1},
$g'_{\mu \nu } = \e^{\pm \kappa \varphi \sqrt{\frac{2}{3}}}
g_{\mu \nu }$,
can cancel the kinetic term of the scalar field in the non-phantom
case only. In order to solve this problem, we here suggest to use a
complex conformal transformation which will lead to a
(generally speaking, complex) modified $F(R)$ gravity.

  From now on, we will restrict our discussion to the phantom scalar.
To start, one can repeat step by step the calculations made in
\cite{art1} by using a complex conformal transformation given by
\be
\label{F2}
g'_{\mu \nu} = \e^{\pm i \kappa \varphi \sqrt{\frac{2}{3}}} g_{\mu \nu}
\ee
Using it in (\ref{F1}), one arrives at
\be
\label{F3}
S = \int d^4 x \, \sqrt{-g} \left\{ \frac{\e^{\pm i \kappa \varphi \sqrt{\frac{2}{3}}} R}{2\kappa^2}
 - \e^{\pm i 2 \kappa \varphi \sqrt{\frac{2}{3} }} V(\varphi)\right\}\ ,
\ee
where the kinetic term of the scalar field  disappears. Now, the
scalar  $\varphi$ is just an auxiliary field and can be expressed
in terms of the scalar curvature as $\varphi=\varphi(R)$, by using
the equation of motion
\be
\label{F4}
R = \e^{\pm i \kappa \varphi \sqrt{\frac{2}{3} }} \left( 4 \kappa^{2}
V(\varphi) \mp i \sqrt{6} \kappa V'(\varphi) \right) \ .
\ee
Hence, the scalar-tensor action appears under the form of $F(R)$
gravity:
\be
\label{F5}
S = \int d^{4}x \sqrt{-g} F(R) \ .
\ee
Here the function $F(R)$ is given by the expression
\be
\label{F6}
F(R) \equiv \e^{\pm i \kappa \varphi \sqrt{\frac{2}{3} }}
\frac{R}{2\kappa^2} - \e^{\pm i 2 \kappa \varphi \sqrt{\frac{2}{3} }} V(\varphi)\ .
\ee
Note that, generally speaking, the curvature and the action itself
can easily become complex, as they may contain a non-zero imaginary
part after performing the complex conformal transformation. This fact
indicates to known physical problems of phantom from another side.

Let us consider the simple example where the potential is given by
\be
\label{FF1}
V(\varphi) = V_0 \e^{ a k \varphi}\ .
\ee
Then, we find
\be
\label{FF2}
R = 2 \kappa^{2} V_0 \left( 2 \mp i \sqrt{\frac{3}{2}} a \right) \e^{ \kappa \varphi \left( R  \right)
\left( a \pm i \sqrt{ \frac{2}{3} } \right) },
\ee
and
\be
\label{FF3}
F(R) = V_0 \left( 1 \mp i a \sqrt{\frac{3}{2}}  \right) \left( \frac{R}{2 \kappa^{2} V_0
\left( 2 \mp i a \sqrt{\frac{3}{2}} \right) }\right)^{\frac{a \pm 2 i \sqrt{\frac{2}{3}} }
{a \pm i \sqrt{\frac{2}{3}} } \left( 1 + 2 n \pi i \right) }\ .
\ee

As is clear from (\ref{F4}), if the scalar field $\varphi$ is real,
the scalar curvature $R$ will not be always real. In order for $R$ to
be real, the following condition should be fulfilled
\be
\label{F7}
\e^{i \kappa \varphi \sqrt{\frac{2}{3} }} \left( 4 \kappa^{2}
V(\varphi) - i \sqrt{6} \kappa V'(\varphi) \right)
=\e^{- i \kappa \varphi \sqrt{\frac{2}{3} }} \left( 4 \kappa^{2}
V(\varphi) + i \sqrt{6} \kappa V'(\varphi) \right) \ ,
\ee
which is satisfied for a potential like
\be
\label{F8}
V(\varphi)=\frac{V_0}{\cos\left(\kappa\varphi\sqrt{\frac{2}{3}}\right)}\ .
\ee

Except for the case (\ref{F8}), the curvature $R$ in  (\ref{F4}) is
not real for real $\varphi$. There still could be, however, the
possibility that, after formally solving (\ref{F4}) with respect to
$\varphi$, if we substitute the expression into (\ref{F3}), the
resulting action might be real. Let us assume that the potential
$V(\varphi)$ contains only even powers of $\varphi$, and write this
potential as $V(\varphi)=U(\varphi^2)$. This is a typical situation
in quantum field theory (note that some properties of the phantom are
indeed similar to those of a QFT, as was shown in \cite{noplb}, while
quantum effects may render an effective phantom cosmology
\cite{woodard}). Since Eq.(\ref{F4}) can be rewritten as
\be
\label{F9}
R = \e^{\pm i \kappa \varphi \sqrt{\frac{2}{3} }} \left( 4 \kappa^{2}
U(\varphi^2) \mp i 2 \sqrt{6} \kappa U'(\varphi^2) \varphi \right) \ ,
\ee
this tells us that $\varphi$ is pure imaginary, if $R$ is real. Then,
from the expression of (\ref{F6}), we find that the action (\ref{F5})
is indeed real. As $\varphi$ is purely imaginary, we can write it as
$\varphi=i\phi$ and use the real field $\phi$. Then, even if we start
with the plus sign in front of the kinetic term of the scalar field,
in (\ref{F1}) ---which corresponds to a phantom--- it turns out that
\be
\label{F10}
S = \int d^4 x \, \sqrt{-g'} \left\{\frac{R'}{2\kappa^2} -
\frac{1}{2} \partial_{\mu} \phi \partial^{\mu} \phi  - U(-\phi^2)\right\} \ ,
\ee
which corresponds to a non-phantom theory. Therefore, even if we
started with a phantom theory, in order that  the $F(R)$ gravity
action could be real, the corresponding scalar-tensor theory reduces
to a non-phantom theory, except for the case (\ref{F8}). We should
note, however, that in the action (\ref{F10}), by analytic
continuation, the potential becomes sometimes negative, as we will
see later. This might be taken as a footprint of the original phantom
nature of the theory. One further remark is in order. Some time ago
complex general relativity attracted considerable interest, for
different reason. In any case, if one starts with complex modified
gravity where the whole imaginary part of the metric may be included
into the conformal factor, then by an inverse transformation, such
complex modified gravity can be mapped into a phantom
scalar-tensor theory.

The next simple example is the model of a massive phantom and a
cosmological constant:
\be
\label{F11}
S = \int d^4 x \, \sqrt{-g'} \left\{\frac{R'}{2\kappa^2}
+ \frac{1}{2} \partial_{\mu} \varphi \partial^{\mu} \varphi
 - \frac{2\alpha^2}{\kappa^2} - 6\alpha^2 \varphi^2
\right\} \ .
\ee
Here $\alpha$ is a constant. A cosmological solution is given by
\be
\label{F12}
a=a_0\e^{\alpha^2 t^2}\ ,\quad \varphi=\frac{2\alpha t}{\kappa}\ .
\ee
Since $\dot H=2\alpha^2>0$, the solution (\ref{F12}) actually expresses
a super-accelerated (phantom) expanding (if $t>0$) universe.
For the potential (\ref{F11}), Eq.(\ref{F4}) tells us that $\varphi$
is purely imaginary. In fact, when we define $\phi=-i\varphi$,
Eq.(\ref{F4}) becomes real
\be
\label{F13}
R=\e^{\mp\kappa\phi}\left(8\alpha^2\left(1 - 3 \kappa^2 \phi^2\right) \pm
12 \sqrt{6} \alpha^2 \kappa \phi\right)\ ,
\ee
and can be solved as $\phi=\phi(R)$, and then we have the following
$F(R)$ theory, from (\ref{F6}),
\be
\label{F14}
F(R)=\frac{\e^{\mp i \kappa \varphi\sqrt{2/3}} R}{2\kappa^2}
 - \frac{2\alpha^2 \e^{\mp 2\sqrt{2/3} i\kappa \phi(R)}
\left(1 - 3\kappa^2 \phi(R)^2\right)}{\kappa^2}\ .
\ee
By inversely transforming the $F(R)$ gravity into the scalar-tensor
theory, with the usual procedure, instead of (\ref{F11}), one obtains
\be
\label{F15}
S = \int d^4 x \, \sqrt{-g'} \left\{\frac{R'}{2\kappa^2}
 - \frac{1}{2} \partial_{\mu} \phi \partial^{\mu} \phi
 - \frac{2\alpha^2}{\kappa^2} + 6\alpha^2 \phi^2 \right\} \ .
\ee
As the mass is negative, the scalar field is a tachyon. The action
does not yield the same solution as (\ref{F12}), but if we Wick-rotate
the time-coordinate $t$ as $t\to i\tau$, we obtain the Euclidean
solution
\be
\label{F16}
a=a_0\e^{-\alpha^2 \tau^2}\ ,\quad \phi=\frac{2\alpha \tau}{\kappa}\ .
\ee
We should point out that, in $F(R)$ gravity, owing to the scale
transformation (\ref{F2}), the metric looks rather different
\be
\label{F16b}
ds_{F(R)}^2 = \e^{\pm i\kappa\varphi \sqrt{\frac{2}{3}}}\left(d\tau^2
+ a(\tau)^2 \sum_{i=1}^3 \left(dx^i\right)^2 \right)
= d\tilde\tau^2 + \frac{2 a_0^2 \alpha^2 \tilde\tau^2}{3}\e^{-\frac{3}{2}
\left(\ln\left(\mp\sqrt{\frac{2}{3}}\alpha \tilde\tau \right) \right)^2}
\sum_{i=1}^3 \left(dx^i\right)^2 \ .
\ee
Here
\be
\label{F16c}
\tilde\tau\equiv \mp \sqrt{\frac{3}{2}}\frac{\e^{\mp\sqrt{\frac{2}{3}}\alpha\tau}}{3}\ .
\ee
The metric (\ref{F16b}) has a conical singularity unless
\be
\label{F16d}
2 a_0^2 \alpha^2 =3\ .
\ee

The third example is
\be
\label{F17}
S = \int d^4 x \, \sqrt{-g'} \left\{\frac{R'}{2\kappa^2}
+ \frac{1}{2} \partial_{\mu} \varphi \partial^{\mu} \varphi
 - V_0\cos\left(\kappa\varphi\sqrt{\frac{2}{3}}\right)\right\} \ .
\ee
By solving (\ref{F4}), we find
\be
\label{F18}
\e^{\pm 2i\kappa\varphi\sqrt{2/3}}=-3 + \frac{R}{V_0\kappa^2}\ ,
\ee
and therefore $\varphi$ is clearly imaginary. The corresponding $F(R)$
gravity (\ref{F6}) is given by
\be
\label{F19}
F(R)=V_0\sqrt{ - 3 + \frac{R}{V_0\kappa^2}}\ .
\ee
Since $\varphi$ is imaginary, by putting $\varphi=-i\phi$,
the action (\ref{F17}) acquires the following form:
\be
\label{F20}
S = \int d^4 x \, \sqrt{-g'} \left\{\frac{R'}{2\kappa^2}
 - \frac{1}{2} \partial_{\mu} \phi \partial^{\mu} \phi
 - V_0\cosh\left(\kappa\phi\sqrt{\frac{2}{3}}\right)\right\} \ ,
\ee
which can be also obtained by inversely transforming the $F(R)$ gravity
(\ref{F19}) into the scalar-tensor theory with the above procedure.
In (\ref{F20}), when $|\phi|$ is large, the action behaves as
\be
\label{F21}
S \sim \int d^4 x \, \sqrt{-g'} \left\{\frac{R'}{2\kappa^2}
 - \frac{1}{2} \partial_{\mu} \phi \partial^{\mu} \phi
 - \frac{V_0}{2}\e^{\kappa|\phi|\sqrt{\frac{2}{3}}}\right\} \ .
\ee
Hence, we have a solution like
\be
\label{F22}
\phi\sim \kappa\sqrt{\frac{2}{3}}\ln \left|\frac{t}{t_1}\right|\ ,\quad
H=\frac{3}{t}\ ,\quad t_1\equiv \frac{48}{\kappa^2 V_0}\ .
\ee
Since from Eq.(\ref{F18}) it follows that
\be
\label{F23}
R=V_0\kappa^2\left(3+\e^{2\kappa\phi\sqrt{\frac{2}{3}}}\right)\ ,
\ee
we find $R\to 3V_0\kappa^2$ or $R\to +\infty$ when $|\phi|\to \infty$.

Thus, we have shown  that a phantom scalar may be always mapped into
complex modified gravity. In cases of an even scalar potential, the
corresponding, equivalent modified gravity is real.

\section{The Big Rip singularity: phantom versus modified gravity}

In the present section we will compare what happens in the phantom
scalar-tensor theory and in modified gravity when a finite-time
singularity (Big Rip) appears in either of these theories.
Let us consider the following example
\be
\label{F24}
S = \int d^4 x \, \sqrt{-g'} \left\{\frac{R'}{2\kappa^2}
+ \frac{1}{2} \partial_{\mu} \varphi \partial^{\mu} \varphi
 - V_0\cosh\left(2\frac{\varphi}{\varphi_0}\right)\right\} \ .
\ee
Since $V_0\cosh\left(2\frac{\varphi}{\varphi_0}\right)
\sim \frac{V_0}{2}\e^{2|\varphi|/\varphi_0}$,
when $\varphi$ is large, we have the following asymptotic solution
\be
\label{F25}
\varphi\sim \varphi_0\ln \left|\frac{t_0 - t}{t_1}\right|\ ,\quad
H\sim \frac{\kappa^2 \varphi_0^2}{4(t_0 - t)}\ ,\quad
t_1^2 \equiv \frac{\varphi_0^2\left( 1 + \frac{3\kappa^2 \varphi_0}{4} \right)}{v_0}\ ,
\ee
which exhibits a Big Rip singularity \cite{brett} at $t=t_0$ (for the
classification of future, finite-time singularities, see
\cite{tsujikawa}). Hence, Eq.(\ref{F4}) yields
\be
\label{F26}
R=\e^{\pm i\kappa\varphi\sqrt{\frac{2}{3}}}\left(4\kappa^2 \cosh\frac{2\varphi}{\varphi_0}
\mp i \frac{2\sqrt{6}\kappa}{\varphi_0}\sinh\frac{2\varphi}{\varphi_0}\right)
=\e^{\mp\kappa\phi\sqrt{\frac{2}{3}}}\left(4\kappa^2 \cos\frac{2\phi}{\varphi_0}
\mp \frac{2\sqrt{6}\kappa}{\varphi_0}\sin\frac{2\phi}{\varphi_0}\right)\ .
\ee
Solving (\ref{F26}) with respect to $i\varphi$ or $\phi$ and using
(\ref{F5}) and (\ref{F6}), we
obtain an $F(R)$ gravity. Since the Big Rip singularity corresponds
to $|\varphi|\to \infty$, the scalar
curvature (\ref{F26}) in $F(R)$ gravity becomes complex and large
as in (\ref{FF2}). In particular, when $\phi\to +\infty$, one finds
\be
\label{F26b}
R\propto \e^{\left(\pm i\kappa\sqrt{\frac{2}{3}} + \frac{2}{\varphi_0}\right)\varphi}\ ,
\ee
which gives a complex $F(R)$ theory with
\be
\label{F26c}
F(R)\propto R^{2 - \frac{\frac{2}{\varphi_0}}{\pm i
\kappa \sqrt{\frac{2}{3}} + \frac{2}{\varphi_0}}}\ .
\ee
Therefore, there is no solution in the corresponding $F(R)$ gravity
which could be also obtained from a non-phantom theory
\be
\label{F27}
S = \int d^4 x \, \sqrt{-g'} \left\{\frac{R'}{2\kappa^2}
 - \frac{1}{2} \partial_{\mu} \phi \partial^{\mu} \phi
 - V_0\cos\left(2\frac{\phi}{\varphi_0}\right)\right\} \ .
\ee
The point corresponding to the Big Rip singularity only appears when
$\phi$ is analytically continued to be imaginary. This clearly
demonstrates the physical non-equivalence between the phantom and the
corresponding modified gravity theories: even when the (phantom)
scalar-tensor theory can be mapped into a real $F(R)$ gravity, the
FRW accelerating solution of the scalar-tensor theory might be mapped
into the corresponding FRW solution in the $F(R)$ theory only
{\it partially}. When the scalar-tensor FRW metric becomes singular
(the Big Rip occurs) the equivalent $F(R)$ gravity becomes complex and
the singularity does not show up. This is a quite general situation
in the examples we have discussed. Nevertheless, one may also expect
that in some specific cases the transformation of the scalar-tensor
theory into modified $F(R)$ gravity may become singular precisely at
the point where the Big Rip occurs. The generic conclusion
is that when the phantom Big Rip occurs, there is no possibility
to transform the Big Rip region of the scalar-tensor theory to a
reliable (real) modified gravity sector.

Conversely, the Big Rip singularity can occur even in $F(R)$ gravity,
for instance, if \cite{art2}
\be
\label{CrEx1}
F(R)=f_0\e^{R/6H_0^2}\ ,
\ee
with constant $f_0$ and $H_0$ \cite{art2}. We now consider what could
occur in the corresponding scalar-tensor theory. Let us rewrite the
general action of $F(R)$ gravity (\ref{F5}) as a scalar-tensor theory.
By introducing the auxiliary fields, $A$ and $B$, one can rewrite
the action (\ref{F5}) as follows
\be
\label{RR2b}
S=\int d^4 x \sqrt{-g} \left[{1 \over
\kappa^2}\left\{B\left(R-A\right) + F(A)\right\} \right]\ .
\ee
Then, one is able to eliminate $B$, to obtain
\be
\label{RR6b}
S=\int d^4 x \sqrt{-g} \left[{1 \over \kappa^2}
\left\{F'(A)\left(R-A\right) + F(A)\right\}\right]\ ,
\ee
and using the conformal transformation
$g_{\mu\nu}\to \e^\sigma g_{\mu\nu}$
$\left(\sigma = -\ln F'(A)\right)$,
the action (\ref{RR6b}) can be rewritten as the Einstein-frame action
\be
\label{RR10}
S_E=\int d^4 x \sqrt{-g} \left[{1 \over \kappa^2}\left( R - {3 \over 2}g^{\rho\sigma}
\partial_\rho \sigma \partial_\sigma \sigma - V(\sigma)\right)
\right]\ .
\ee
Here,
\be
\label{RR11b}
V(\sigma) = \e^\sigma G\left(\e^{-\sigma}\right) -
\e^{2\sigma} f\left(G\left(\e^{-\sigma} \right)\right)
= {A \over F'(A)} - {F(A) \over F'(A)^2}\ .
\ee
The action  (\ref{RR6b}) is called the Jordan-frame action (a
recent comparison of the equivalence between the Einstein and the
Jordan frames can be found in \cite{faraoni}). If we identify
$\varphi=\sqrt{3}\sigma/\kappa$, we obtain the action of the
scalar-tensor theory (\ref{F1}). The action thus obtained is not the
phantom one, since the scalar field in (\ref{RR11b}) has a canonical
kinetic term. Near the Big Rip singularity, the scalar curvature
in $F(R)$ gravity becomes large and then, for the model (\ref{CrEx1}),
we find $\sigma\propto - R/6H_0^2$. As a consequence, $\sigma$ becomes
negative and large, and the potential $V(\sigma)\to 0$.

When $F(R)$ behaves as $F(R)\sim R^{-n}$, the scalar factor behaves as
\be
\label{F28}
a\sim \left(t_0 - t\right)^{\frac{(n+1)(2n+1)}{n+2}}\ .
\ee
Thus, when $n<-2$ or $-1<n<-1/2$,  a singularity of the
Big Rip type can appear at $t=t_0$. In this case, we find
\be
\label{F29}
\sigma\sim (n+1)\ln R \sim -2(n+1)\ln (t_0 - t)\ ,
\ee
since
\be
\label{F30}
R\sim \frac{6(n+1)(2n+1)(4n+5)n}{(n+2)^2(t_0 - t)^2}\ .
\ee
Then, in the corresponding scalar-tensor theory,
the time coordinate $\tilde t$ could be
given by $d\tilde t=\pm \e^{\sigma}dt\sim \pm (t_0 - t)^{n+1}dt$,
that is $\tilde t=\pm (t_0-t)^{n+2}$.
Therefore, when $n<-2$, $t\to t_0$ corresponds to $t\to \pm \infty$.
As a consequence, the singularity changes its structure: it does
not appear in finite time for the scalar-tensor theory.
On the other hand, when $n>-2$, $t\to t_0$ corresponds to $t\to 0$.
We also find that the metric in the scalar-tensor theory behaves as
\be
\label{F31}
ds_{ST}^2=\e^{\sigma} \left(-dt^2 + a(t)^2\sum_{i=1,2,3}(dx^i)^2\right)
\sim -d\tilde t^2 + \tilde a(t)^2 \sum_{i=1,2,3}(dx^i)^2\ ,\quad
\tilde a(t)^2 \sim a_0^2 \tilde t^{\frac{(n+1)(2n+5)}{(n+2)^2}}\ ,
\ee
with a constant $a_0$. Since in the case $-1<n<-1/2$,
 $a(t)^2\to 0$ when $\tilde t\to 0$, the Big Rip singularity could be
replaced with a Big Crunch, where the universe shrinks indefinitely.

Hence, the Big Rip singularity which may occur in some versions of
$F(R)$ gravity may change its structure in the equivalent,
scalar-tensor theory. Either it  goes to the infinite past or future
($n<-2$), or the Big Rip singularity is replaced with a Big Crunch
singularity ($-1<n<-1/2$) in the corresponding scalar-tensor theory.
Thus, generically, there is in fact a {\it mathematical} equivalence
between the phantom scalar-tensor theory and the corresponding modified
gravity, the Big Rip singularity region being then the part of the
solution where {\it physical} equivalence is lost. This may be due to
the actual non-existence of one of the corresponding theories
precisely in this  region, or either to a total change of the
structure and properties of the singularity.

\section{The FRW instanton with a spatially non-flat metric
both in the phantom theory and in modified gravity}

In Sect.~2 we have studied the analytic continuation of the phantom
scalar field, while in \cite{reconstruction}, the reconstruction
scenario for the scalar-tensor theory was considered. In this
formulation the scalar field is identified with the time coordinate.
As we analytically continue the scalar field to pure imaginary values
(Sect.~2), the time coordinate could be also analytically continued.
Then one could obtain a kind of an instanton solution. In
\cite{reconstruction}, only the case that the spatial part is flat was
considered, since the spatial part of the observed universe is
approximately flat. In order to obtain a finite action for the
instanton solution, however, we may consider the case when the spatial
part is spherical. But even if the spatial part is flat or a
hyperboloid, dividing the manifold by using a discrete group, we can
also obtain a finite action. In either way, we are able to extend the
formulation of \cite{reconstruction} to the (phantom) case when the
spatial part is not flat
\be
\label{FFF1}
ds^2=-dt^2 + a(t)^2 d\Omega^2 \ .
\ee
Here $d\Omega^2$ is the metric of either the three-dimensional flat
space, the hyperboloid, or the sphere with unit radius. The action of
the scalar-tensor theory is chosen to be
\be
\label{FFF2}
S = \int d^4 x \, \sqrt{-g} \left\{\frac{R}{2\kappa^2}
 - \frac{1}{2} \omega(\varphi) \partial_{\mu} \phi \partial^{\mu} \phi  - V(\phi)\right\} \ ,
\ee
By assuming that the scalar field $\varphi$ only depends on time, the
FRW equations give
\be
\label{FFF3}
0=- \frac{3}{\kappa^2} + \frac{1}{2}\omega^\phi {\dot\phi}^2 + V(\phi) - \frac{3k}{2\kappa^2 a^2}\ ,\quad
0=\frac{1}{\kappa^2} \left(\dot H + 3 H^2\right) + \frac{1}{2}\omega^\phi {\dot\phi}^2 - V(\phi)
+ \frac{k}{2\kappa^2 a^2}\ .
\ee
If $d\Omega^2$ in (\ref{FFF1}) is the metric of the sphere, we
have $k=2$, if it corresponds to a hyperboloid, $k=-2$, and in the
flat case $k=0$. Since there is freedom in redefining the scalar field
$\phi$, we may choose $\phi=t$. Then, we obtain
\be
\label{FFF4}
\omega(\phi)=-\frac{2}{\kappa^2}\dot H + \frac{k}{\kappa^2 a^2}\ ,\quad
V(\phi)=\frac{1}{\kappa^2}\left(\dot H + 3H^2 \right) + \frac{k}{\kappa^2 a^2}\ .
\ee
As a consequence, if we consider the model where $\omega(\phi)$ and
$V(\phi)$ are given by
\be
\label{FFF5}
\omega(\phi)=-\frac{2}{\kappa^2}g''(\phi) + \frac{k\e^{-2g(\phi)}}{\kappa^2 a_0^2}\ ,\quad
V(\phi)=\frac{1}{\kappa^2}\left(g''(\phi) + 3g'(\phi)^2 \right) + \frac{k\e^{-2g(\phi)}}{\kappa^2 a_0^2}\ ,
\ee
there is the following solution
\be
\label{FFF6}
\phi=t\ ,\quad H=g'(t)\ \left(a=a_0\e^{g(t)}\right).
\ee

First we consider a special (and trivial) example with $k=2$ (sphere):
\be
\label{FFF7}
g(\phi)=\ln \cosh\frac{\phi}{a_0}\ .
\ee
Then by using (\ref{FFF6}), we find that the metric is given by
\be
\label{FFF8}
ds^2 = - dt^2 + a_0\cosh \frac{t}{a_0}d\Omega^2\ ,
\ee
which is nothing but the metric of deSitter space. In fact,
(\ref{FFF4}) yields
\be
\label{FFF9}
\omega(\phi)=0\ ,\quad V(\phi)=\frac{3}{a_0^2\kappa^2}\ .
\ee
Therefore, the scalar field $\phi$ does not appear in this action
(\ref{FFF2}), there only appears the cosmological term, where the
cosmological constant is given by $V$ in (\ref{FFF9}). As well-known,
by Wick-rotating the time coordinate $t$ by $t\to ia_0\tau$, the metric
 (\ref{FFF8}) is transformed into the metric of the sphere with radius
$a_0$:
\be
\label{FFF10}
ds^2 = a_0^2\left(d\tau^2 + \cos^2\tau d\Omega^2\right)\ .
\ee
The solution (\ref{FFF10}) can be regarded as an instanton.

Let us consider the second non-trivial example, again when $k=2$:
\be
\label{FFF11}
g(t)=\ln\left\{\frac{1}{2}\left(1 + \frac{t^2}{a_0^2}\right)\right\}\ ,
\ee
which gives
\be
\label{FFF12}
\omega(\phi)=\frac{4}{\kappa^2 a_0^2}\left( 1 + \frac{\phi^2}{a_0^2}\right)^{-2}\ ,\quad
V(\phi)=\frac{10}{a_0\kappa^2}\left(1 + \frac{\phi^2}{a_0^2}\right)^{-1}\ .
\ee
Eq.(\ref{FFF11}) tells us that
\be
\label{FFF13}
a(t)=\frac{a_0}{2}\left( 1 + \frac{t^2}{a_0^2}\right)^2\ .
\ee
We may analyticaly continue the scalar field $\phi$ as $\phi=i\rho$,
which gives
\be
\label{FFF14}
\omega(\rho)=\frac{4}{\kappa^2 a_0^2}\left( 1 - \frac{\rho^2}{a_0^2}\right)^{-2}\ ,\quad
V(\rho)=\frac{10}{a_0\kappa^2}\left(1 - \frac{\rho^2}{a_0^2}\right)^{-1}\ .
\ee
In the Euclidean signature, $t=i\tau$, the scalar-tensor theory leads
to the following metric, instead of (\ref{FFF13}),
\be
\label{FFF15}
ds^2=d\tau^2 + \frac{a_0}{2}\left( 1 - \frac{\tau^2}{a_0^2}\right)^2d\Omega^2\ .
\ee
This metric seems to be singular when $\tau\to \pm a_0$. Since the
metric behaves as
\be
\label{FFF16}
ds^2 \sim d\tau^2 + \left(\tau - a_0\right)^2 d\Omega^2\ ,
\ee
when $\tau \to -a_0$ and
\be
\label{FFF17}
ds^2 \sim d\tau^2 + \left(a_0 - \tau\right)^2 d\Omega^2\ ,
\ee
when $\tau \to a_0$, there is no conical singularity when
$\tau\to \pm a_0$. Therefore, this metric (\ref{FFF15}) corresponds
to the instanton.

By changing the scalar field $\phi$ to $\tilde\phi$ as
\be
\label{FFF18}
\phi = a_0 \tan\left(\frac{\kappa}{2}\tilde\phi\right)\ ,
\ee
Eqs.(\ref{FFF14}) yield the following action
\be
\label{FFF19}
S = \int d^4 x \, \sqrt{-g} \left\{\frac{R}{2\kappa^2}
 - \frac{1}{2} \partial_{\mu} \tilde\phi \partial^{\mu} \tilde\phi
 - \frac{10}{a_0^2\kappa^2}\cos^2\left(\frac{\kappa}{2}\tilde\phi\right) \right\} \ ,
\ee
and by analytic continuation of $\tilde\phi$ as $\tilde\phi=i\varphi$,
the action is transformed into
\be
\label{FFF20}
S = \int d^4 x \, \sqrt{-g} \left\{\frac{R}{2\kappa^2}
+ \frac{1}{2} \partial_{\mu} \varphi \partial^{\mu} \varphi
 - \frac{10}{a_0^2\kappa^2}\cosh^2\left(\frac{\kappa}{2}\varphi\right) \right\} \ ,
\ee
which corresponds to a phantom field. Since, from (\ref{FFF18})
\be
\label{FFF21}
\rho = a_0 \tanh\left(\frac{\kappa}{2}\varphi\right)\ ,
\ee
the action (\ref{FFF20}) gives the metric in (\ref{FFF15}). If we
start with the action (\ref{FFF12}), Eqs.(\ref{F4}) and (\ref{F6})
lead to a corresponding $F(R)$ gravity theory. Even if we start with
the action (\ref{FFF19}), we obtain the same $F(R)$ gravity action.
More explicitly, Eq.(\ref{F4}) gives
\be
\label{FFF22}
R=\frac{20\e^{\pm\kappa\tilde\phi\sqrt{\frac{2}{3}}}}{a_0^2}
\left( 2 \cos\left(\kappa\tilde\phi\right) + 2 \mp \sqrt{\frac{3}{2}}\sin\left(\kappa\tilde\phi\right)\right)\ .
\ee
In $F(R)$ gravity, the Lorentz-signature metric has the following form:
\be
\label{FFF23}
ds_{F(R)}^2 = \e^{\mp i\sqrt{\frac{2}{3}} \ln \left(\frac{a_0 + it}{a_0 - it}\right)}
\left( -dt^2 + \frac{a_0}{2}\left( 1 + \frac{t^2}{a_0^2}\right)^2d\Omega^2\right)\ ,
\ee
which seems difficult to continue analytically in the Euclidean
signature.

All the above shows that the properties of the Euclidean-signature
solution in
$F(R)$ gravity are quite different from those in the corresponding
(phantom) scalar-tensor theory. Therefore, this indicates again a
certain physical non-equivalence of $F(R)$ gravity as compared to the
corresponding (phantom) scalar-tensor theory. Note also that the
conclusions of the previous section about the Big Rip singularity do
not change for the case of the spatially non-flat FRW universe.

\section{Discussion}

In summary, we have here demonstrated that any phantom scalar theory
can be mapped into a mathematically equivalent one, which is a
complex modified gravity. The
fact that the mathematically-equivalent theory is complex can be
taken, generically speaking, as an indication of some problems
(already well known in fact) concerning the physical properties
of the phantom field.

For even scalar potentials, the ensuing modified gravity turns
out to be real. Nevertheless, even in this case it becomes complex
in the region where the scale factor develops the well-known Big Rip
singularity. Thus, the correspondence we have unveiled helps a lot to
clarify the origin of the Big Rip singularity.

  From a different perspective, we have also seen that, when some
version of
$F(R)$ gravity develops an effectively phantom universe, with a
possible future Big Rip, the corresponding scalar-tensor theory is
not a phantom one. Moreover, the structure of the Big Rip in the
modified gravity changes to some other type of singularity in the
scalar-tensor theory: it usually becomes an infinite-time one or
has a totally different nature. Even a transmutation from a Big Rip
to a Big Crunch singularity is possible.

As a final remark, let us recall the known fact that any
(canonical or phantom) scalar-tensor theory can be presented, in an
equivalent form, as a fluid obeying some equation of state. This
equivalence can then easily be extended to modified gravity
\cite{art1}. As a result, using our connection here, an ideal fluid
phantom dark energy model might be also presented as modified $F(R)$
gravity, and the comparison of the properties of both theories is to
be performed in a way similar to what we have done in this work.

\section*{Acknowledgements}

This investigation was completed during a stay of FB in Barcelona.
It has been supported in part by
MEC (Spain), projects BFM2003-00620 and PR2006-0145, by JSPS (Japan)
XXI century COE program of Nagoya University project 15COEEG01, by
Monbu-Kagaku-sho grant no.18549001 (Japan), by LRSS project
N4489.2006.02, by RFBR grant 06-01-00609 (Russia), and by AGAUR
(Gene\-ra\-litat de Catalunya), contract 2005SGR-00790.

\end{document}